# Microwave Hearing Effect: Cochlea as a Demodulator for Pulsed Microwaves

Sandeep Narasapura Ramesh and Om P Gandhi, *Member, IEEE*

*Abstract*—The phenomenon of Microwave Hearing Effect (MHE) can be explained by assuming the inner ear, specifically the Human Cochlea to act as a receiving antenna for pulsed Microwaves. The spiral structure of the Cochlea picks up these incoming waves and demodulates it due to it's directional conductivity and a net voltage is induced which explains the audible 'clicks' as observed in MHE. Further, the maximum Electromagnetic absorption is observed near the side of the head where the cochlea is located.

## I. INTRODUCTION

WHEN A person's head is exposed to a strong(but sub thermal) level of modulated or pulsed microwaves, sound is heard, often as clicks or a buzz depending on the pulse rate. This effect is termed as Microwave Hearing Effect and was first observed by military personnel working in close distances near RADAR facilities. The effect was first observed and brought to notice of the scientific community by a neuroscientist Allan Frey and hence it's also known as Frey effect. [1]

Of the competing theories explaining the mechanism of Microwave Hearing Effect, some prominent researchers have favored the thermo-elastic stress model[2], in which sudden heating in the order of $10^{-6}$ degrees C, cause brain tissue to expand suddenly, resulting in an acoustic wave to be heard at the ear. Disagreements have remained, however, as the theory did not explain why some researchers could not detect the normal "microphonic" electrical signals from the cochlea that would be expected (microphonics were later seen, but only under somewhat different conditions), nor could it explain why a 2in square metal shielding could block the effect, but only if it was placed in front of the ear.

The study performed in the paper explains the phenomenon of MHE in a new light. We hypothesize the spiral structure of Cochlea to act as a receiving antenna which picks up the pulsed Microwave radiation and demodulates because of its directionally dependent conductivity. The net voltage induced as result of the demodulation is transmitted to the auditory nerve and further processed as audible clicks.

## II. HUMAN HEAD MODEL

### A. MIDA Model

The Multimodal Imaging-Based Detailed Anatomical Model of Human Head and Neck (MIDA) is a computer model which has an excellent spatial resolution of 0.5 mm and a very good representation of all the internal structures. The model is developed using high quality Magnetic Resonance Imaging techniques and several raw images of the human head are obtained using slices in the steps of 500 microns [5]. The number of tissue classifications in the model is 116 (Cochlea being one of them). Also, it's availability in .vox, .raw and .mat formats makes it very convenient to import into Computational Electromagnetic solvers and also custom FDTD codes.

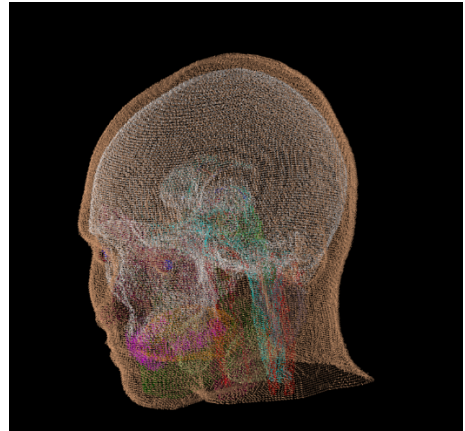

Fig 1: MIDA Human Head Model

### B. Dielectric Tissue Properties

The comprehensive database of dielectric properties for biological tissues as investigated by *C.Gabriel et al* is used for our study. Out of the 116 tissues identified in the MIDA model, we use the significant 30 out of them. The human Skull, Gray Matter, White Matter, Cerebellum, Skin, Blood, Cerebro Spinal Fluid and the Spinal Cord holds more than 80 percent of our model. The smaller muscles, nerves, tongue, teeth etc. are also included which increases the complexity but have a small but finite effect on the results obtained.

The dielectric properties i.e. Relative permittivity and conductivity are input for various frequencies used in the study and curve fitted. The Figure 2 represents the dispersive dielectric properties for Brain Gray Matter. Similar curves are obtained for all the other tissues used in the study.

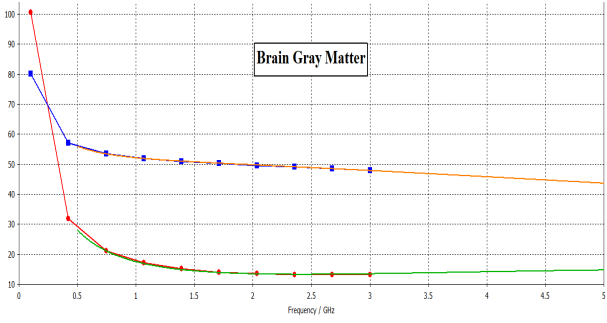

Fig 2. Dispersive Dielectric Properties of Brain Gray Matter

### C. Boundary Conditions

To reduce the infinite universe to a finite space because of the lack of computational resources we use boundary conditions to simplify our problem. The Perfectly Matched layer (PML) as developed by *Berenger et al* [3]is implemented in our study. PML is an absorbing artificial boundary conditions which absorbs all the incident radiation with a very low reflection coefficient and a finite computational domain is made possible.

Further the MIDA model is truncated to reduce the number of grid cells and computation time. One of the PML walls is placed at 5 cm beyond the Cochlea (into the human head). The skin depth at our lowest frequency of interest is still around 1 cm, so it is safe to assume that the brain tissue absorbs all the radiation at a length of 5 skin depths. Similar truncation is implemented with PML at vertical and horizontal directions as shown in the figure below.

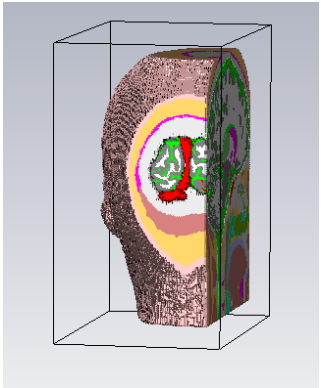

Figure 3: MIDA model after truncation using PML

### D. XFDTD

A powerful Electromagnetic simulation software XFDTD developed by Remcom is used in our study. It used the Finite Domain Time Difference method to find approximate solutions to Maxwell's Wave Equations and can find results for a wide range of frequencies from a single run. The salient features of XFDTD which made us choose it over other EM solvers are as follows[6]:

- It has a built in GPU accelerator which can

significantly reduce the simulation time
- Parallel computing technology which uses more than one processor for simulations
- Built in Near to Far Field transformation algorithm which in our case is used to calculate Gain
- Built in PML implementation to reduce the complexity of our problem
- MATLAB export functionality to postprocess the obtained results
- Interactive Graphics which can help us to visualize the MIDA model and results
- Large Database of Dielectric properties of Biological tissues.

## III. FORWARD AND REVERSE METHOD

We solve our problem using two different techniques which we term them as Forward and Reverse Methods.

In the Forward problem, we incident the Pulsed Microwave radiation onto the Human head and compute the Electric Field concentration at the stapes region of the human cochlea. Further, using the number of voxels in the stapes region we can calculate the voltage induced at this point. This is defined to be our forward problem.

In the Reverse problem, we place a voltage source in the stapes region of the Cochlea. We excite the source in a sinusoidal fashion at the frequency of interest and using the Near to Far Field Algorithm compute the Far field Gain. Using the equation for open circuit voltage from Antenna Theory as mentioned below, Input impedance and an assumed Incident power density we can calculate the voltage induced at the stapes region of the Cochlea.

$$V_{oc} = \sqrt{\frac{8 \; S|_{inc} \lambda^2 \; G \; R_a}{4 \; \pi}}$$

In the above equation:
1. $S|inc$ is the Incident Power Density
2. G is the far field realized Gain
3. $R_a$ is the Input Impedance at the source

According to the reciprocity theorem in Antenna Theory which is based on time reversity and conservation of energy, the voltages computed using both the Forward and Reverse techniques should be the same. We will revisit this method several times in the following sections of the paper.

## IV. VALIDATION TESTS

Before we dive in to calculating results from our head model, we run a series of validation tests on Half wave Dipole Antenna, Uni Filar Spiral Antenna and Bi Filar Spiral Antenna. The motivation behind running these tests are as follows:

- The results of these problems are standard and can

be obtained from Antenna Theory makes us trust the results from the head model and increases the credibility of XFDTD solver whose algorithms we have no access to

- This also gives us the chance to prove the validity of the open circuit voltage which will be used in our study.

The validation tests are performed on a half wave dipole with a total length of 0.14 meters and a Uni filar and Bifilar Spiral Antenna in Air with dimensions corresponding to that of human cochlea. (i.e. inner radius of 1mm and outer radius of 5mm).

Presented below in Table 1 are the validation results for the reverse problem:

| Test Run | Frequency | Realized Gain | Input Resistance | $V_{oc}$ |
|---|---|---|---|---|
| Half Wave Dipole | 1 GHz | 1.76 | 67.76 Ω | 0.0951 V |
| Unifiliar Spiral | 25 GHz | 2.64 | 118.468 Ω | 0.0062 V |
| Bifilar Spiral | 25 GHz | 3.76 | 147.43 Ω | 0.0082 V |

Table 1: Validation tests for the Reverse problem

Please note that the unifiliar and bifilar spiral antennas with cochlear antennas are placed in air for the above runs. This explains the radiating frequency of 25 GHz. The frequencies are scaled down when the antennas are placed inside human head because of dielectric loading.

Presented below in Table 2 are the validation test results for the forward problem:

| Test Run | Frequency | $V_{oc}$ |
|---|---|---|
| Half Wave Dipole | 1 GHz | 0.0926 V |
| Unifiliar Spiral | 25 GHz | 0.0058 V |
| Bifilar Spiral | 25 GHz | 0.0081 V |

Table 1: Validation tests for the Forward problem

The results from Table 1 and Table 2 are in excellent agreement which increases our confidence in the commercial solver and the $V_{oc}$ equation from Antenna Theory.

Further we run the tests using Unifiliar and Bifilar Spiral Antennas with different conductivities. Since we will be dealing with biological tissues in our study which are dispersive and heterogenous, it's crucial that we assume different material properties in the validation tests.

The inner and outer radii of the Unifiliar Antenna correspond to the cochlear dimension of 1mm and 5mm. The upper and lower cutoff frequencies are calculated using the formula:

$$f_{low} = \frac{c}{2\pi r_{outer}}$$

$$f_{high} = \frac{c}{2\pi r_{inner}}$$

- c is the speed of light in the medium which further depends on effective permittivity
- $f_{low}$ and $f_{high}$ are lower and upper cutoff frequencies
- $r_{outer}$ and $r_{inner}$ are inner and outer radii of the Spiral.

The magnitude lower and upper cutoff frequencies as determined from the above table correspond to the frequencies of 10 GHz and 50 GHz.

Presented below in Table 3 are the validation results for the reverse problem for a Unifiliar Spiral Antenna with varying conductivities at a frequency of 25 GHz:

| Dielectric Properties | Gain | $R_{in}$ | $V_{oc}$ |
|---|---|---|---|
| 1000 S and $\varepsilon_r$ = 45 | 3.0756 | 32.543 Ω | 0.0034893 V |
| 100 S and $\varepsilon_r$ = 45 | 1.6032 | 28.915 Ω | 0.0023747 V |
| 10 S and $\varepsilon_r$ = 45 | 1.2823 | 30.469 Ω | 0.0021801 |

Table 3: Validation tests for the Reverse problem for a Unifiliar with different dielectric properties.

Presented below in Table 4 are the validation results for the forward problem for a Unifiliar Spiral Antenna with varying conductivities and circularly polarized plane wave a frequency of 25 GHz incident on the Antenna:

| Dielectric Properties | $V_{oc}$ |
|---|---|
| 1000 S and $\varepsilon_r$ = 45 | 0.00347 V |
| 100 S and $\varepsilon_r$ = 45 | 0.002357 V |
| 10 S and $\varepsilon_r$ = 45 | 0.002 V |

Table 4: Validation tests for the Forward problem for a Unifiliar with different dielectric properties.

The above tables 4 and 5 are in excellent agreement which gives us the confidence that our commercial solver can handle materials with different dielectric properties and the $V_{oc}$ equation from Antenna Theory can be applied to problems with different dielectric properties.

Further this analysis was extended to the case of a Bifilar

Spiral and the results are presented in Table 6 and 7 for both the forward and reverse problems. All the parameters i.e. dimension of the Spiral and frequency remain the same.

| Dielectric Properties | Gain | $R_{in}$ | $V_{oc}$ |
|---|---|---|---|
| 1000 S and $\varepsilon_r$ = 45 | 3.25 | 153.40 Ω | 0.0078 V |
| 100 S and $\varepsilon_r$ = 45 | 1.5593 | 133.15 Ω | 0.00507 V |
| 10 S and $\varepsilon_r$ = 45 | 0.5675 | 95.5 Ω | 0.0026 V |

Table 5: Validation tests for the Reverse problem for a Bifilar with different dielectric properties.

| Dielectric Properties | $V_{oc}$ |
|---|---|
| 1000 S and $\varepsilon_r$ = 45 | 0.0076 V |
| 100 S and $\varepsilon_r$ = 45 | 0.0050 V |
| 10 S and $\varepsilon_r$ = 45 | 0.00225 V |

Table 5: Validation tests for the Forward problem for a Bifilar with different dielectric properties.

The results remain in agreement and we now end our validation tests and move to our main problem.

## V. MICROWAVE HEARING EFFECT

### A. Setting up the Model

The Multimodal Imaging-Based Detailed Anatomical Model of Human Head and Neck (MIDA) is used in our studies. As mentioned before it offers a highly detailed representation of the anatomy and has over 100 tissues classified [5]. Also, an isotropic resolution of 500 microns is crucial in our studies since the cochlea has a dimension of 3cm*3cm*0.5cm. This allows for the Cochlea in our model o be distinct and well resolved with a few hundred voxels. Figure 4 shows the Cochlea from the MIDA model. As one can infer the stapes region comprises of 2 voxels where the Voltage source can be placed in the forward problem and a high load can be placed in the reverse problem.

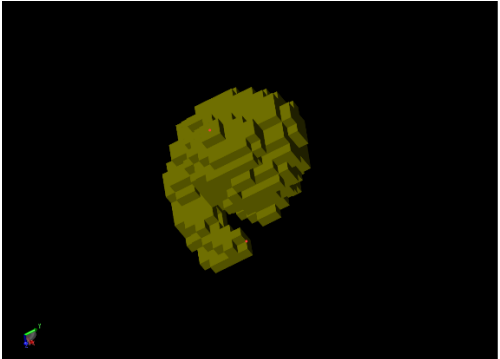

Figure 4: Cochlea in the MIDA model

The model is truncated with a PML layer on all the 8 sides and dispersive dielectric properties are attributed to 30 significant tissues in the model. The example of one of the tissues is shown in Figure 2. This model is now imported to XFDTD to run Finite Domain Time Difference algorithm and MATLAB is used to postprocess results obtained from different runs.

### B. Simulations

The first simulation that was run using the MIDA model was at a frequency of 2.5 GHz where the maximum absorption of EM is observed and the effects of Microwave Hearing is felt at the highest according to the other research sources.

We assume different conductivities for the Cochlea and solve both our forward and reverse problem. We summarize our results in the Table 6 below:

| Dielectric Properties | Gain | $R_{in}$ | $V_{oc}$ |
|---|---|---|---|
| 1000 S and $\varepsilon_r$ = 45 | 0.0142 | 15.556 Ω | 0.0016 V |
| 100 S and $\varepsilon_r$ = 45 | 0.0021 | 18.921 Ω | 0.00069 V |
| 10 S and $\varepsilon_r$ = 45 | 0.000150 | 29.41 Ω | 0.00021 V |
| 5 S and $\varepsilon_r$ = 45 | 0.000151 | 23.061 Ω | 0.000205V |

Table 6: MIDA model results for the Reverse problem at 2.5 GHz with different dielectric properties assumed for Cochlea.

The corresponding Forward problem is now investigated by a Circularly Polarized Plane wave of frequency 2.5 GHz incident at the MIDA model. The results are presented in Table 6 below:

| Dielectric Properties | $V_{oc}$ |
|---|---|
| 1000 S and $\varepsilon_r$ = 45 | 0.00107 V |
| 100 S and $\varepsilon_r$ = 45 | 0.00044 V |
| 10 S and $\varepsilon_r$ = 45 | 0.001853 V |
| 5 S and $\varepsilon_r$ = 45 | 0.0015 V |

Table 7: MIDA model results for the Forward problem at 2.5 GHz with different dielectric properties assumed for Cochlea.

By comparing the Voltages from Table 6 and Table 7 we see more than 70 percent agreement in the values obtained and most importantly the trend remains almost the same.

The second run which is most important to study MHE is done by assuming a dielectric value for the Cochlea and looking at the Voltages at various frequencies. We assume a 5 S and $\varepsilon_r$ = 45 material property for the Cochlea and calculate voltages from 0.5 GHz to 5 GHz.

This is represented in Table 8 below for the Forward problem:

| Frequency | Gain | $R_{in}$ | $V_{oc}$ |
|-----------|------|----------|----------|
| 0.5 GHz | $2.27e^{-06}$ | 186.331 Ω | 0.0003588 V |
| 1 GHz | 0.2979 | 0.0884 Ω | 0.0014 V |
| 2 GHz | $9.0012e^{-05}$ | 33.238 Ω | 0.0002384 V |
| 2.5 GHz | 0.000151 | 23.061 Ω | 0.000205 V |
| 5 GHz | 0.0007345 | 7.241 Ω | 0.000127 V |

Table 8: MIDA model results for the Reverse problem at various frequencies with dielectric properties of 5 S and $\varepsilon_r$ = 45 assumed for Cochlea.

We now use the same model and incident circularly polarized plane waves of the same frequency set and observe the Voltages in Table 9.

| Frequencies | $V_{oc}$ |
|-------------|----------|
| 0.5 GHz | 0.000248 V |
| 1 GHz | 0.00143 V |
| 2 GHz | 0.00019 V |
| 2.5 GHz | 0.00015 V |
| 5 GHz | 0.0000818 V |

Table 8: MIDA model results for the Forward problem at various frequencies with dielectric properties of 5 S and $\varepsilon_r$ = 45 assumed for Cochlea.

We can infer from the previous two tables that contrary to other research articles we observe a maximum voltage induced at 1 GHz and the Voltages are validated to an agreeable extent.

To conclude our inference that Cochlea receives the Incident Radiation (i.e. acts as a receiving antenna), we performed an extra run with a linearly polarized plane wave and a Circularly polarized plane wave which has a rotation in the same direction as Cochlea. We just must use the Forward Problem in this case.

| Incident Plane Wave Polarization | $V_{oc}$ |
|----------------------------------|----------|
| Circular polarization (C.P) | 0.00365 V |
| Linear polarization (L.P) | 0.0028 V |

Table 9: MIDA model results for the Forward problem with incident plane wave at different polarizations.

The above results are in excellent agreement with Antenna Theory and the values are in a ratio of the polarization mismatch. (i.e. Spiral picking a L.P v/s C.P)

### C. Inference

From the above results we can deduce these inferences:
- The voltages induced decreases with the decrease in the conductivity of Cochlea. This leads us to believe that if the Cochlea has a directionally dependent conductivity. (i.e. for example, clockwise conductivity 10x the anticlockwise) the incoming plane wave will be rectified and demodulated to induce an effective voltage.
- The maximum voltage induced is at 1 GHz, hence we can conclude that MHE effects are felt the highest at this frequency.
- The Cochlea acting as a receiving antenna is proved by the polarization mismatch results.
- 

## VI. FUTURE WORK

Our hypothesis of directionally dependent conductivity should be proved with experiments in the laboratory. Unfortunately, we could not perform this due to the lack of funding. But this would effectively validate our hypothesis if proved. Also, the dielectric properties of human cochlea has to be determined through experiments for us to model the phenomenon. We intend to perform this work in the future.

## VII. CONCLUSION

We used the Finite Difference Time Domain Method and an anatomical based computer human head model to prove our directional conductivity hypothesis to explain the mechanism for Microwave Hearing Effect. We achieved the intended results using both our predefined Forward and Reverse methods. Also, the validation tests which were run prior to our study were successful and further adds trust to our obtained results.

## VIII. ACKNOWLEDGEMENT

We thank DARPA for funding the initial phase of our project without which we could not have started. We sincerely thank Doug from the IT department at University of Utah for his timely arrangement of the computational resources and maintenance. Dr Cynthia Furse and Dr Lazzi for their valuable suggestions and inputs.

REFERENCES

[1] Frey, Allan H. (July 1962). "Human auditory system response to modulated electromagnetic energy". *Journal of Applied Physiology*. **17** (4): 689–692. doi:10.1152/jappl.1962.17.4.689. PMID 13895081

[2] Lin, JC., "*Microwave auditory effect- a comparison of some possible transduction mechanisms*". J Microwave Power. 1976 Mar;11(1):77–81.

[3] *J. Berenger (1994). "A perfectly matched layer for the absorption of electromagnetic waves". Journal of Computational Physics. 114 (2): 185–200. Bibcode:1994JCoPh.114..185B. doi:10.1006/jcph.1994.1159.*.

[4] Hearing of microwave pulses by humans and animals: effects, mechanism, and thresholds. Lin JC1, Wang Z

[5] MIDA: A Multimodal Imaging-Based Detailed Anatomical Model of the Human Head and Neck . Maria Ida Iacono, Esra Neufeld, Esther Akinnagbe, Kelsey Bower, Johanna Wolf, Ioannis Vogiatzis Oikonomidis, Deepika Sharma, Bryn Lloyd, Bertram J. Wilm, Michael Wyss, Klaas P. Pruessmann, Andras Jakab, Nikos Makris, Ethan D. Cohen, Niels Kuster, Wolfgang Kainz, Leonardo M. Angelone.

[6] REMCOM - HHM REMCOM - Inc., State College, PA, 2016